# Growth of Silver Nanoclusters Embedded in Soda Glass Matrix

P. Gangopadhyay[#], P. Magudapathy, R. Kesavamoorthy, B. K. Panigrahi, K. G. M. Nair and P. V. Satyam[*]

Materials Science Division, Indira Gandhi Centre for Atomic Research, Kalpakkam-603102, India

[*]Institute of Physics, Bhubaneswar 751 005, India

**Abstract**

Temperature-controlled-growth of silver nanoclusters in soda glass matrix is investigated by low-frequency Raman scattering spectroscopy. Growth of the nanoclusters is ascribed to the diffusion-controlled precipitation of silver atoms due to annealing the silver-exchanged soda glass samples. For the first time, Rutherford backscattering measurements performed in this system to find out activation energy for the diffusion of silver ions in the glass matrix. Activation energy for the diffusion of silver ions in the glass matrix estimated from different experimental results is found to be consistent.

---

[#] Electronic mail: pganguly@igcar.ernet.in, Fax:+91-4114-280081

# 1. Introduction

Nanoscale materials are currently under active research because they possess interesting physical properties differing considerably from that of the bulk phase. Silicate glasses containing metallic clusters of few nanometers in sizes, for instance, are attractive materials for all-optical switching applications in integrated-optical devices due to the enhancement of third-order nonlinear susceptibility ($\chi^{(3)}$) near the surface-plasmon resonance frequency with a fast optical response time (~ few picoseconds)[1]. Precise control of cluster sizes during the synthesis of metal nanoclusters in a glass matrix is of prime importance as the optoelectronic properties of nanocluster materials depend primarily on the physical dimensions of metal clusters in the glass[2,3]. To achieve a better control on the sample preparation, a basic understanding of the mechanisms governing formation and evolution of metal nanoclusters while processing the novel materials is of great interest.

Heterogeneous materials consisting of metallic nanoclusters in amorphous matrices are synthesized by a variety of methods. Low energy ion-beam mixing[4], sol-gel[5], direct metal-ion implantation[6,7], light ion irradiation of ion-exchanged glasses[8], annealing of ion-exchanged glasses in hydrogen atmosphere[9], evaporation-condensation[10] are, for example, among a few successful methods to prepare composites of this nature. Ion-exchange in soda glasses followed by light-ion irradiation is now widely accepted technique to modify the linear and nonlinear optical properties of glasses, mainly because they are easy to prepare and commercially viable for technological applications in optoelectronic devices. Recently, in different context, we have reported interesting charge transport phenomena due to hopping between isolated nanoislands of embedded silver in such ion-exchanged glasses[11]. However, in this article, we plan to elucidate the growth behavior of silver nanoclusters and kinetics of silver accumulation near the glass surface due to thermal annealing the silver-exchanged soda glass samples in a high-vacuum atmosphere. While the optical techniques (Raman and absorption spectroscopy) are insensitive regarding any depth information about the precipitation process, Rutherford backscattering spectrometry helps us to illustrate the accumulation of silver near the glass surface as the sample temperature is increased gradually. Moreover, through suitable analysis of Rutherford backscattering data, activation energy for the diffusion of silver ions in the soda glass matrix is estimated. This finding has not yet been presented by backscattering technique in this system, to best of our knowledge.

**2. Experimental details**

Soda glass was chosen as the dielectric medium for embedding the silver nanoparticles. The composition of the glass used in the present experiment is Si-21.49%, Na-7.1%, Ca-5.78%, Mg-0.34% and Al-0.15%, by weight. Silver-exchanged soda glass samples were prepared by immersing the preheated soda glass slides (10 x 10 x 1 mm) for 2 min in a molten salt bath of $AgNO_3$ and $NaNO_3$ (1:4 weight ratio) mixture at 320 $^o$C. The obtained silver-exchanged glass samples were annealed for 1 hr in vacuum at a pressure lower than $1 \times 10^{-6}$ mbar at different temperatures between 200 and 600$^o$C.

Silver concentration profiles in these samples, before and after thermal annealing, were measured by Rutherford backscattering spectrometry (RBS). An analyzing beam of $^4He^+$ at an energy of 2 MeV, backscattered by an angle of 160$^o$, was used during the backscattering measurements. Optical absorption measurements were performed on silver-exchanged soda glass samples before and after annealing using a dual-beam spectrophotometer (Shimadzu PC 3101) in 200-1000 nm wavelength range. In order to determine the size of the silver nanoclusters by direct means, high-resolution transmission electron microscopy (JEOL-2000, EX-II, operated at 140 KV) was used. For this measurement, sample was prepared in the following way: a diamond scriber (HR 100A, Karl Suss, Germany) was utilized to scratch the surface (and a few micron below) of the glass containing silver nanoclusters. Minute powders thus obtained were carefully dropped in methanol and the medium was agitated ultrasonically for about 30 mins. Due to the agitation, larger glass particles settled below and the finer ones floated on the surface of the liquid. Subsequently, the finest layer on the surface was collected on a carbon coated copper grid for the observation. The Low-frequency Raman scattering measurements were carried out on these samples in the backscattering geometry. Vertically polarized 488 nm line of an argon ion laser (Coherent, USA) with 60 mW power was used to excite the samples. Polarized or depolarized scattered light from the sample was dispersed using a double monochromator (Spex, model 14018) and detected using a cooled photomultiplier tube (FW ITT 130) operated in the photon counting mode. Scanning of the spectra and data acquisition were carried out using a microprocessor based data-acquisition-cum-control system. The Raman spectra were recorded from 5 cm$^{-1}$ to 40 cm$^{-1}$ at 0.5 cm$^{-1}$ step with 10 s integration time. The instrument resolution was 1.4 cm$^{-1}$.

### 3. Results and discussions

After annealing (in vacuum) the silver-exchanged soda glass samples at 320$^o$C, characteristic surface-plasmon resonance (SPR) absorption of silver nanoclusters around 420 nm has started developing (as seen in figure 1). The resonance optical absorption is due to collective oscillation of the conduction electrons with respect to the ionic background in nanoscale silver clusters. Peak position of the optical response (SPR) of the matrix-embedded silver nanoclusters is observed to be insensitive to temperature variation (figure 1). Intensity of absorption at 420 nm increases significantly while the width of the absorption band reduces systematically with the increase of annealing temperature, indicating an increase of volume fraction of silver nanoclusters in the glass matrix. Later, we will discuss this point further with the experimental results of RBS measurements.

Low-frequency Raman scattering measurements were carried out in the silver-exchanged glass samples before and after annealing to estimate an average size of silver nanoclusters in the glass matrix. Confined surface acoustic phonons in metal nanoclusters give rise to low frequency Raman modes in the vibrational spectra of the materials. The vibrational modes of a spherical or an ellipsoidal particle are characterized by two indices, **l** and **n**, where **l** is the angular momentum quantum number and **n** (**n**=0 represents the surface modes) is the branch number. The quadrupolar modes (**l**=2, eigenfrequency $\eta_2^s$) of surface acoustic vibrations (phonons) appear in both polarized and depolarized geometry, whereas the surface symmetrical modes (**l**=0, eigenfrequency $\xi_o^s$) appear only in the polarized geometry[12]. The polarized low-frequency Raman spectra are displayed in figure 2 for silver nanoclusters with different sizes and distributions in the glass matrix after annealing at indicated temperatures. In the as-exchanged glass sample, the vibrational spectrum exhibits a broad (FWHM ~ 8 cm$^{-1}$) Raman peak at about 20 cm$^{-1}$ (bottommost graph of figure 2(a)). The Raman peak appeared in both polarized and depolarized geometry. Symmetrical surface mode ($l=0$) of acoustic vibrations is expected to appear only in the polarized Raman spectra. Hence, the observed mode of surface acoustic vibrations is assigned to the quadrupolar ($l=2$) Raman mode. The intensity of the mode in depolarized Raman spectra was very small and hence, polarized spectra were presented and analyzed.

For nanoscale clusters (assumed to be homogeneous elastic sphere with a free surface), nondimensionalized eigenfrequencies of spheroidal modes of confined surface acoustic vibrations are given by[13],

$$\eta_2^s = \frac{\omega_2^s \mathbf{d}}{2\mathbf{v_t}}, \quad \xi_o^S = \frac{\omega_o^S \mathbf{d}}{2\mathbf{v_l}}$$

In case of silver nanoclusters, Raman peak frequency corresponding to quadrupolar surface acoustic vibration modes (for which $\eta_2^s$ is calculated to be $\approx 2.65$) can be expressed as,

$$v_2^s = 0.85 \frac{\mathbf{v_t}}{\mathbf{dc}} \quad (1)$$

where $\mathbf{v_t}$ (1660 m/sec) is the transverse sound velocity (averaged over different directions) in silver, $\mathbf{d}$ is an average dimension of silver nanoclusters and $\mathbf{c}$ is the velocity of light in vacuum. Low-frequency Raman scattering data were fitted to an exponential background (with a dotted line) and a Lorentzian line-shape function (continuous line) to estimate the Raman peak frequency. The systematic shift of the Raman peak to lower frequencies with the increase of annealing temperature (as seen in figure 2) indicates the continual growth (since $v_2^s \propto \mathbf{d}^{-1}$) of the embedded silver nanoclusters in the glass matrix. Inset to figure 2(b) shows the electron micrograph of the silver nanoclusters for the 600 °C annealed sample. Average size of the silver nanoclusters (10 nm), obtained from the micrograph agrees well with the size obtained from low-frequency Raman measurements.

Explanation for the growth of the silver nanoclusters in the present case can be provided following thermodynamic arguments. Nucleation of silver atoms can take place due to the fluctuations in the local concentrations of diffusing silver ions/atoms (reactants) on the onset of the silver-sodium ion-exchange process since silver is known to form a supersaturated solid solution due to its low solubility in the glassy medium. For the growth of the nucleated clusters, one would expect it to grow by the diffusion of reactants to the surface of the stable nuclei. In the regime of diffusion-controlled-growth of spherical clusters, average size of the cluster increases with time t as[14,15],

$$d^2(t) = d_0^2 + 8\frac{C_s - C_e}{C_p - C_e}Dt \qquad (2)$$

where $d_0$ is the value of $d$ at $t = 0$ (which is the average size of clusters prior to growth), $C_p$ is the concentration of limiting reactant in the cluster and $D$ is the diffusion coefficient for the limiting reactant. $C_s$ and $C_e$ are the concentration of limiting reactant prior to growth and the equilibrium concentration in the matrix, respectively. The degree of supersaturation, $C_s - C_e$ decreases during this stage of growth. In the later stage of growth process when degree of supersaturation becomes negligible, coarsening (where larger clusters continue to grow at the loss of smaller ones) takes place. In the present experimental studies, we might be still on the early stage of growth because of the possible diffusion and reduction processes of available silver ions. This supply of $C_s$ would maintain the degree of supersaturation in the system and this could also prohibit coarsening to be effective within the time of annealing treatments of 1 hr. Growth of the silver nanoclusters measured here are related to the diffusion-controlled processes. Particularly, low-frequency Raman scattering measurements have shown a systematic growth of silver nanoclusters with the increase of annealing (for a fixed time of 1 hr) temperature (figure 2). The basic relation (equation 2) for the diffusion controlled growth has been applied to the Raman scattering data here to estimate the thermal growth of the silver nanoclusters in the glass matrix. The Arhenius plot of $\ln(d^2 - d_0^2)$ vs. $T^{-1}$ is drawn in figure 3 to calculate activation energy for the diffusion of silver in the matrix. The slope of the fitted line directly gives an estimation of activation energy and this works out to be about 0.72 eV. Activation energy for the diffusion of silver atoms in the soda glass matrix is about 0.28 eV which is rather low compared to the values obtained here. Since thermal diffusion of silver ions is expected to occur with higher activation energy[15], it is more than likely that the diffusing species here are silver ions.

Rutherford backscattering spectra for the silver-exchanged glass are shown in figure 4. For the annealed samples, partial spectra (high energy channels) are shown here to emphasize the backscattering from silver only. A complete spectrum, shown in the inset of figure 4 for the silver-exchanged soda glass sample (prior to annealing), reveals other elements (Ca, Si, Na and O) present in the glass. The backscattering data were analyzed with the help of RUMP software program[16] for calculating the depth distribution of silver atoms/ions inside the silver-exchanged glass samples. It is observed from the backscattering spectra that silver accumulated near the glass surface during annealing. Accumulation of silver is higher for the samples annealed at higher temperatures (figure 4). Near-surface accumulation is due to the thermal diffusion of silver ions in the silver-exchanged glass. This outward diffusion of silver ions relaxes the stress (which arises due to size difference of $Ag^+$ and $Na^+$) and minimizes the total energy in the system.

Putting together the information obtained through the analysis of different experimental results from the present studies, it can be inferred that the growth of the silver nanoclusters might be taking place within a depth of ~ 100 nm below the glass surface. Eventhough the clusters of silver atoms may be present up to few microns from the surface, the growth seems to be predominant near the surface of the glass, as revealed by RBS investigations.

De Marchi et al[9] have reported backscattering measurements in silver-exchanged glasses after annealing in hydrogen atmosphere. But, no attempt was made to analyze backscattering data to estimate an activation energy for the diffusion of silver in these studies. In this work we have estimated the activation energy of silver diffusion by suitable analysis of RBS data. A depth scale of about 100 nm below the glass surface over which the silver got accumulated during annealing was estimated from RBS data. This depth is significantly small compared to the depth (~5000 nm) to which silver has diffused during the silver ion-exchange process. This short-range diffusion of silver during annealing the silver-exchanged glass allows us to consider the case as a semi-infinite system with the bulk acting as a constant source of known density, say $c_0$. Applying diffusion theory for a semi-infinite system, accumulated mass of silver per unit area ($m$) on the surface can be calculated as[17,18], $m = c_0\sqrt{Dt/\pi}$ where $t$ is the time of diffusion in the sample held at temperature $T$; $D$ is the diffusion coefficient of silver ions at temperature $T$. Assuming $D$ to vary as $e^{-\frac{\varepsilon_a}{kT}}$ ($\varepsilon_a$ is the activation energy for the diffusion of silver ions and $k$ is the Boltzmann constant), $m$ can be rewritten as,

$$\ln(m) = K_1 - \frac{\varepsilon_a}{2kT} \qquad (3)$$

where $K_1$ is another constant under the present experimental conditions. For the samples annealed at different temperatures, accumulated mass of silver per unit area can be estimated using the expression $m = N_A M / N_{Avo.}$ [19] where $N_A$ (number of Ag atoms/ions per cm$^2$) is directly obtained from the backscattering data (using RUMP), $M$ is the atomic weight of silver and $N_{Avo.}$ is the Avogadro number (6.023x10$^{23}$). The Arhenius plot is reported in figure 5 using the equation (3). From the slope of the fitted line in figure 5, thermal activation energy for the diffusion of silver ions in soda glass is estimated to be about 0.74 eV. This agrees well within the limit of experimental inaccuracy with the estimated value obtained here through low-frequency Raman scattering measurements. The activation energy obtained in these experiments is much higher than the activation energy of diffusion of atomic silver in the glass matrix[15]. Hence, we conclude that thermal diffusion of silver ions was predominant in the present study. Since no reducing agents were present during high temperature annealing, the silver ions generated from Ag-O bond breaking could not be neutralized immediately and hence primarily these ions start diffusing through the glass matrix at high temperature.

## 4. Conclusion

Low-frequency Raman scattering measurements were carried out as a function of annealing temperature to characterize thermal growth of silver nanoclusters in the soda glass matrix. The growth was attributed to thermal diffusion of silver ions in the glass matrix. Applying diffusion theory to the Rutherford backscattering data in this system, activation energy for the diffusion of silver ions in the soda glass matrix was estimated.

## Acknowledgments

It is our great pleasure to acknowledge colleagues of Metallurgy and Materials Group, M. Premila, B. Sundaravel, R. Divakar and Rita Saha for their kind help and valuable discussions.

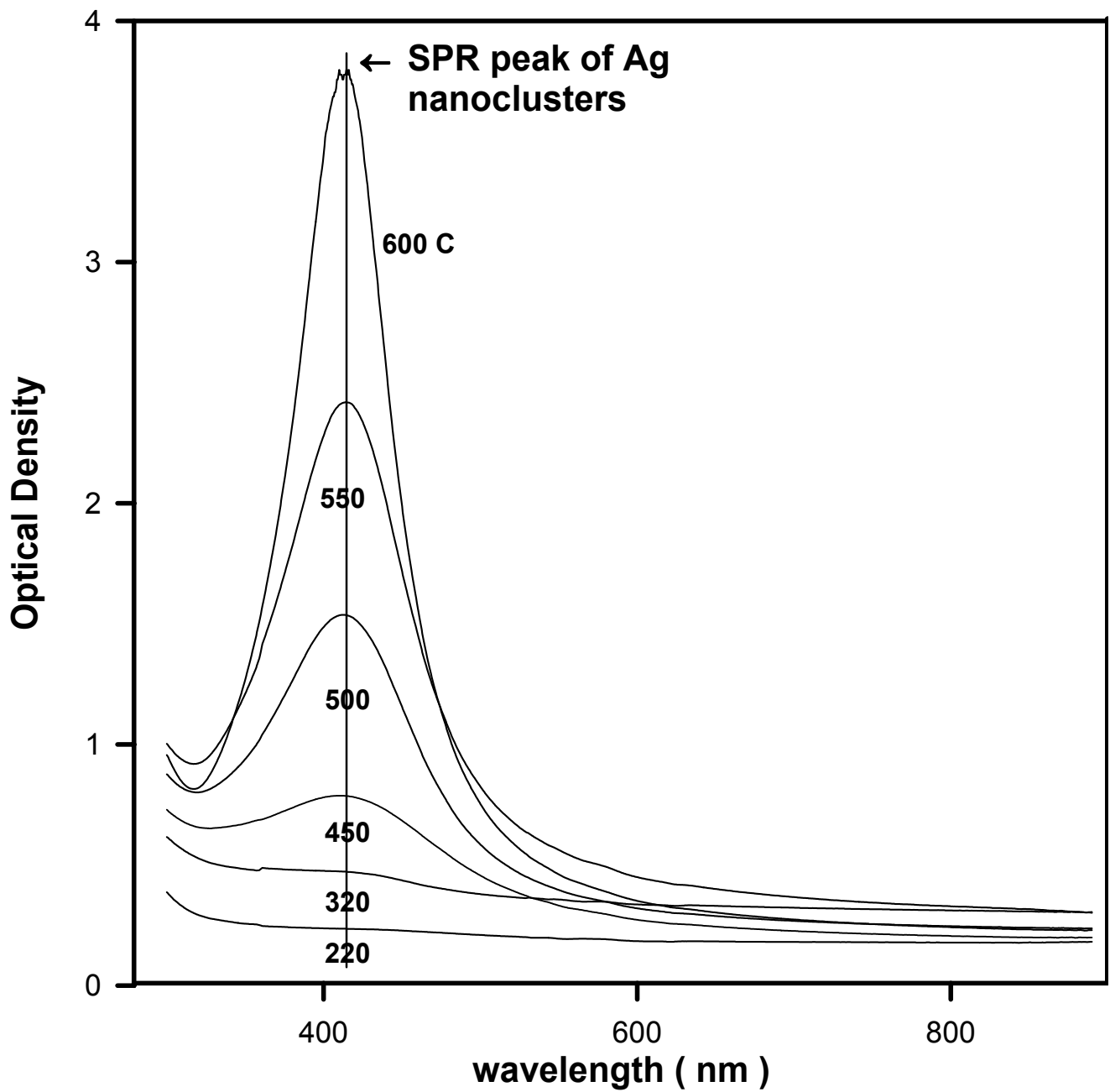

Figure 1. Optical absorption spectra of silver nanoclusters embedded in silver-exchanged soda glass matrix after annealing for 1 hr at different temperatures.

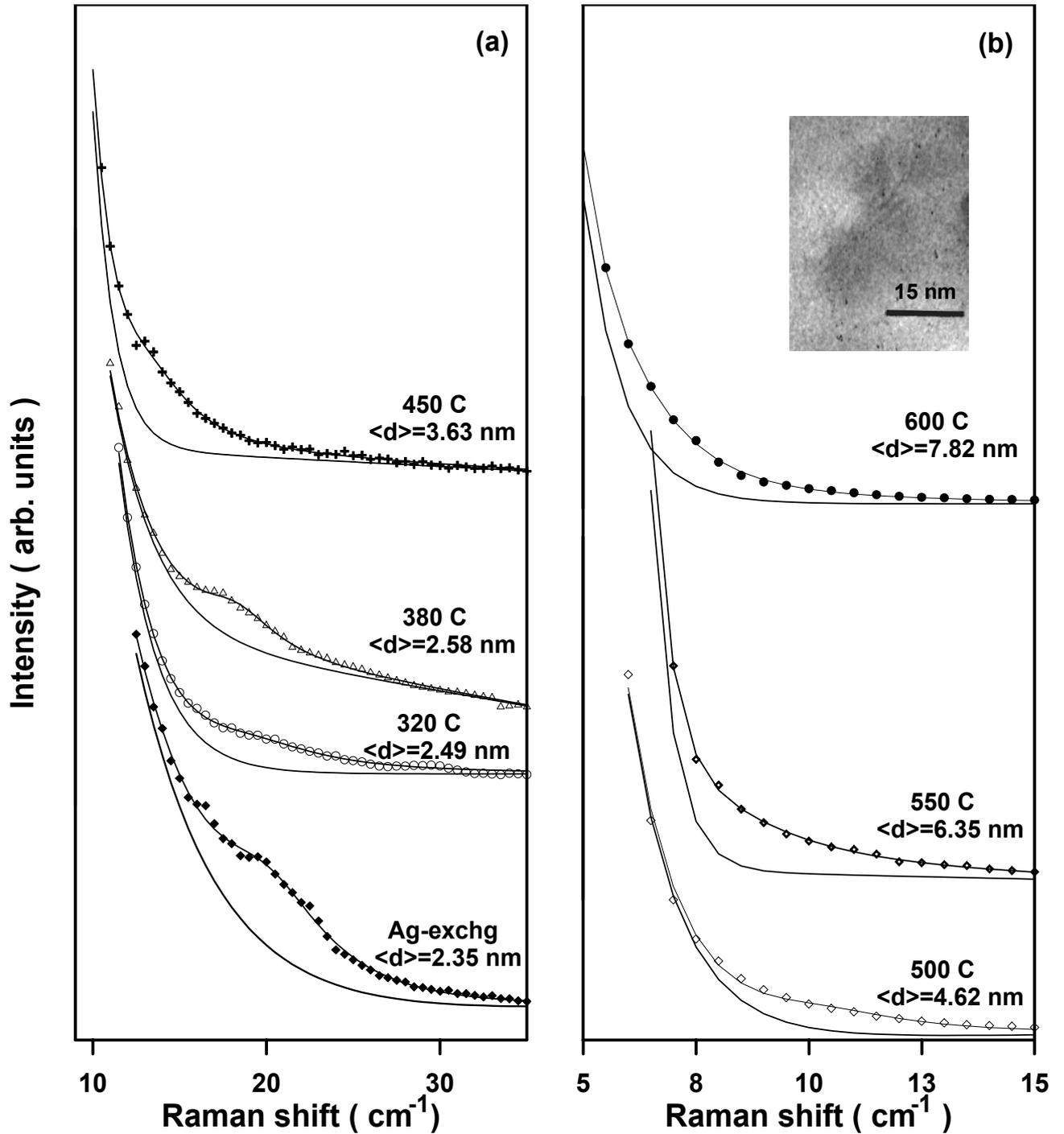

Figure 2. Polarized Raman spectra of silver nanoclusters in silver-exchanged soda glass. Each spectrum is shifted along the intensity axis for clarity. Symbol is the data, broken line is the exponential background and solid line is the fit to the data with the background and Lorentzian lineshape function. Inset figure in (b) shows the images of silver nanoclusters in silver-exchanged soda glass after annealing at 600 °C.

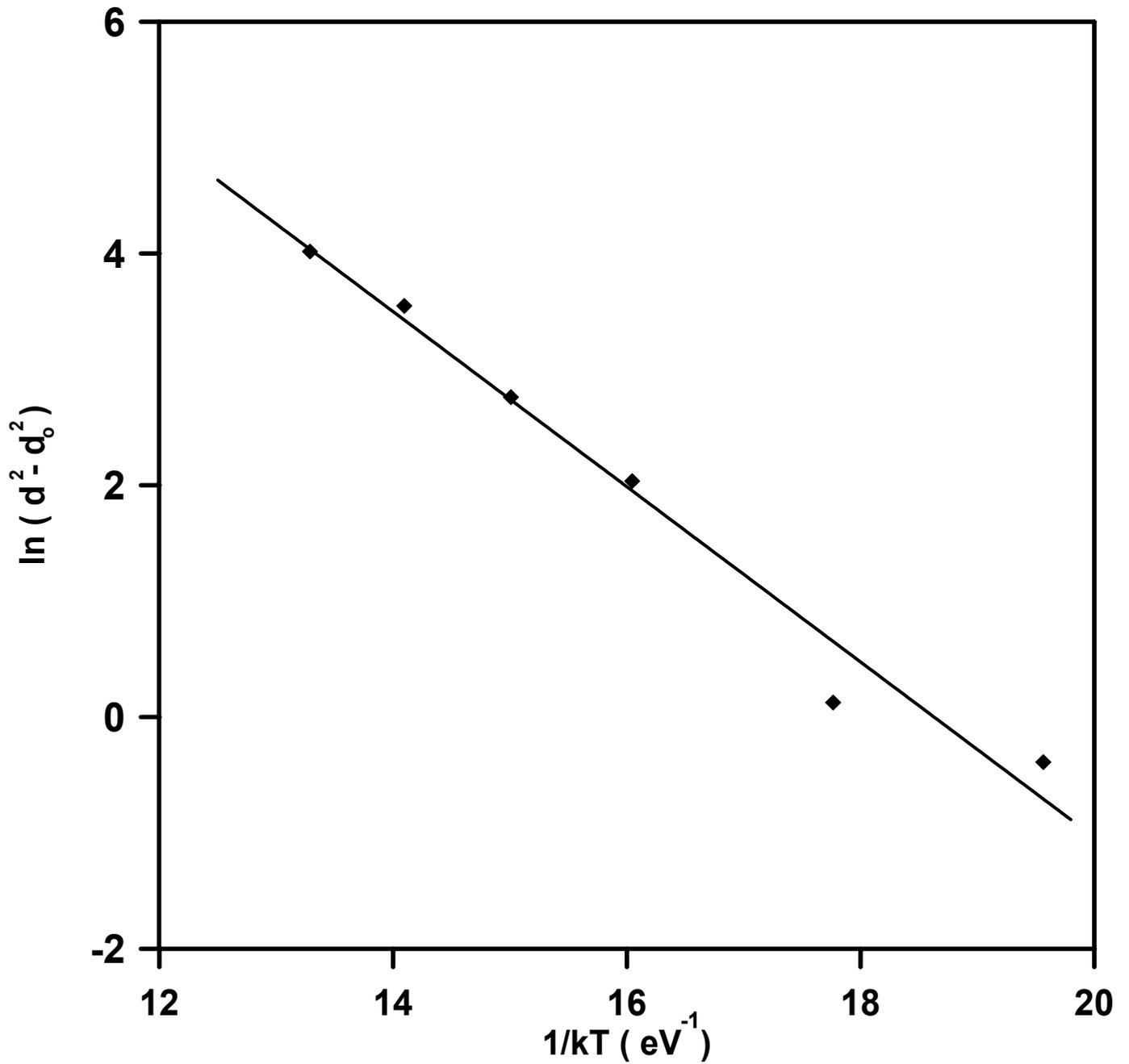

Figure 3. Arhenius plot to estimate the activation energy for the diffusion of silver ions in a soda glass matrix. Solid line through the points is the linear fit.

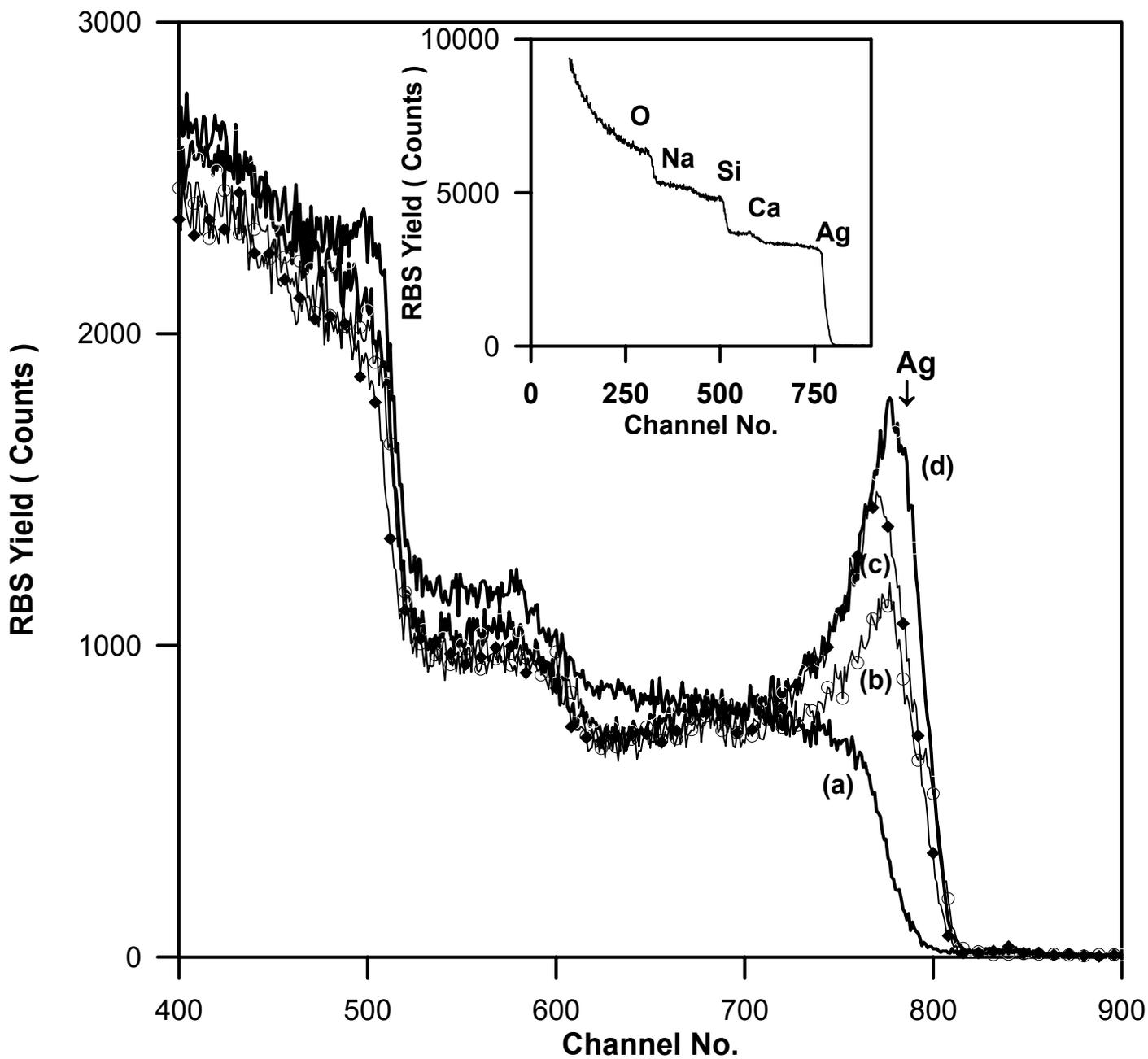

Figure 4. Rutherford backscattering spectra for the silver-exchanged soda glass after annealing for 1 hr at (a) 320, (b) 450, (c) 550 and (d) 600° C. Inset figure shows the complete spectrum for the silver-exchanged glass sample prior to annealing.

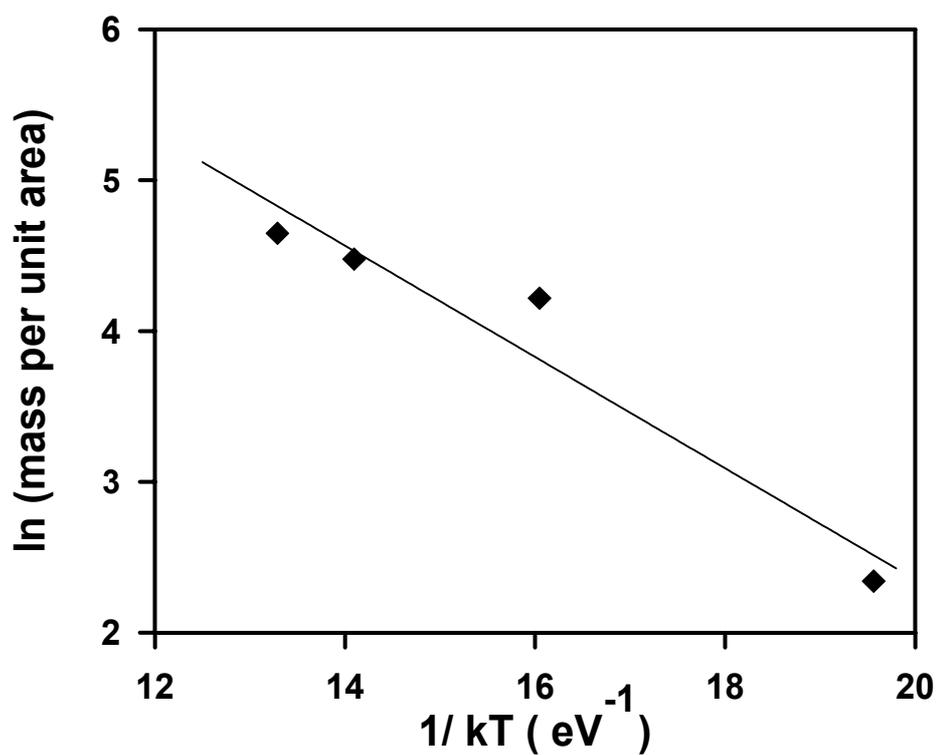

Figure 5. Temperature dependence of mass of silver per unit area near the glass surface shows an Arhenius behavior. Rutherford backscattering data (♦), solid line is the linear fit to the analyzed data. Thermal activation energy for the diffusion of silver ions in a soda glass matrix is estimated as twice the slope of the fitted line.